%% file: paper.tex
\documentclass[aps,prl,nofootinbib,twocolumn,
               superscriptaddress,floatfix,10pt]{revtex4-1}
\pdfoutput=1
\usepackage{amsmath,amssymb}
\usepackage{color,xcolor}
\definecolor{blue}{rgb}{0,0,0.5}
\usepackage[colorlinks,linkcolor=blue,urlcolor=blue,citecolor=blue]{hyperref}
\usepackage{graphicx}
\usepackage{xspace}
\usepackage{relsize}
\usepackage{bm}
\usepackage[utf8]{inputenc}
\usepackage{tabularx,colortbl}

\allowdisplaybreaks

\makeatletter

    \def\CT@@do@color{%
      \global\let\CT@do@color\relax
            \@tempdima\wd\z@
            \advance\@tempdima\@tempdimb
            \advance\@tempdima\@tempdimc
    \advance\@tempdimb\tabcolsep
    \advance\@tempdimc\tabcolsep
    \advance\@tempdima2\tabcolsep
            \kern-\@tempdimb
            \leaders\vrule
                    \hskip\@tempdima\@plus  1fill
            \kern-\@tempdimc
            \hskip-\wd\z@ \@plus -1fill }
    \makeatother

\begin{document}

\title{Interpreting Hints for Lepton Flavor Universality Violation}

\author{Wolfgang Altmannshofer}
\email{altmanwg@ucmail.uc.edu}
\affiliation{Department of Physics, University of Cincinnati, Cincinnati, Ohio 45221, USA}

\author{Peter Stangl}
\email{peter.stangl@tum.de}
\affiliation{Excellence Cluster Universe, Boltzmannstra{\ss}e~2, 85748~Garching, Germany}

\author{David M. Straub}
\email{david.straub@tum.de}
\affiliation{Excellence Cluster Universe, Boltzmannstra{\ss}e~2, 85748~Garching, Germany}

\begin{abstract}\noindent
We interpret the recent hints for lepton flavor universality violation
in rare $B$ meson decays. Based on a model-independent effective
Hamiltonian approach, we determine regions of new physics parameter
space that give a good description of the experimental data on $R_K$
and $R_{K^*}$, which is in tension with Standard Model predictions.
We suggest further measurements that can help narrowing down viable
new physics explanations.
We stress that the measured values of $R_K$ and $R_{K^*}$ are fully
compatible with new physics explanations of other anomalies in rare
$B$ meson decays based on the $b \to s \mu\mu$ transition.
If the hints for lepton flavor universality violation are first signs
of new physics, perturbative unitarity implies new phenomena below a
scale of $\sim 100$~TeV.
\end{abstract}

\maketitle

\paragraph{Introduction.}
%
The wealth of data on rare leptonic and semi-leptonic $b$ hadron decays that has
been accumulated at the LHC so far allows the Standard
Model (SM) Cabibbo-Kobayashi-Maskawa picture of flavor and CP violation to be tested with
unprecedented sensitivity.
Interestingly, current data on rare $b \to s \ell \ell$ decays show an
intriguing pattern of deviations from the SM predictions both for branching
ratios~\cite{Aaij:2014pli,Aaij:2016flj,Aaij:2015esa} and angular
distributions~\cite{Aaij:2015oid,Wehle:2016yoi}. The latest global fits find
that the data consistently points with high significance to a non-standard
effect that can be described by a four fermion contact interaction $C_9 \,(\bar
s \gamma^\nu P_L b)(\bar \mu \gamma_\nu \mu)$~\cite{Altmannshofer:2017fio} (see
also earlier
studies~\cite{Altmannshofer:2014rta,Descotes-Genon:2015uva,Hurth:2016fbr}).
Right now the main obstacle towards conclusively establishing a beyond-SM
effect is our inability to exclude large hadronic effects as the origin of the
apparent
discrepancies (see
e.g.~\cite{Jager:2014rwa,Descotes-Genon:2014uoa,
Lyon:2014hpa,Ciuchini:2015qxb,Capdevila:2017ert,Chobanova:2017ghn,Blake:2017wjz}).

In this respect, observables in $b \to s \ell \ell$ transitions that are
practically free of hadronic uncertainties are of particular interest. Among
them are lepton flavor universality (LFU) ratios, i.e. ratios of
branching ratios involving different lepton flavors such as~\cite{Hiller:2003js,Bobeth:2007dw,Hiller:2014ula}
\begin{equation}
 R_K = \frac{\mathcal B(B \to K \mu^+\mu^-)}{\mathcal B(B \to K e^+e^-)}
\,,~
R_{K^*} = \frac{\mathcal B(B \to K^* \mu^+\mu^-)}{\mathcal B(B \to K^* e^+e^-)} \,.
\end{equation}
In the SM, the only sources of lepton flavor universality violation are the
leptonic Yukawa couplings, which are responsible for both the
charged lepton masses and their interactions with the Higgs.\footnote{Neutrino
masses provide another source of lepton flavor non-universality, but the effects are
negligible here.}
Higgs interactions do not lead to any observable effects in rare $b$ decays
and lepton mass effects become relevant only for a very small di-lepton
invariant mass squared close to the kinematic limit $q^2 \sim 4m_\ell^2$.
Over a very broad range of $q^2$ the SM accurately predicts $R_K = R_{K^*} = 1$,
with theoretical uncertainties of $O(1\%)$~\cite{Bordone:2016gaq}.
Deviations from the SM predictions can be expected in
various models of new physics (NP), e.g.\ $Z^\prime$ models based on
gauged $L_\mu - L_\tau$~\cite{Altmannshofer:2014cfa,Crivellin:2015mga,Altmannshofer:2015mqa,Crivellin:2016ejn}
or other gauged flavor symmetries~\cite{Crivellin:2015lwa,Celis:2015ara,Falkowski:2015zwa,Bhatia:2017tgo,Ko:2017lzd},
models with partial compositeness~\cite{Niehoff:2015bfa,Carmona:2015ena,Megias:2016bde,GarciaGarcia:2016nvr},
and models with leptoquarks~\cite{Hiller:2014yaa,Gripaios:2014tna,Varzielas:2015iva,Bauer:2015knc,Becirevic:2016yqi,Becirevic:2016oho,Barbieri:2016las,Cox:2016epl,Crivellin:2017zlb}.

A first measurement of $R_K$ by the LHCb collaboration~\cite{Aaij:2014ora} in
the di-lepton invariant mass region $1\,\text{GeV}^2 < q^2 < 6\,\text{GeV}^2$,
\begin{equation}
R_{K}^{[1,6]} = 0.745 ^{+0.090}_{-0.074} \pm 0.036 \,,
\end{equation}
shows a $2.6\sigma$ deviation from the SM prediction.
Very recently, LHCb presented first results for $R_{K^*}$ \cite{LHCbtalk,Bifani:2017gyn,Aaij:2017vbb},
\begin{align}
 R_{K^*}^{[0.045, 1.1]} &= 0.66 ^{+0.11}_{-0.07} \pm 0.03\,,\\
 R_{K^*}^{[1.1, 6]} &= 0.69 ^{+0.11}_{-0.07} \pm 0.05\,,
\end{align}
where the superscript indicates the di-lepton invariant mass bin in GeV$^2$.
These measurements are in tension with the SM
at the level of $2.4$ and $2.5\sigma$, respectively.
Intriguingly, they are in good agreement with the recent $R_{K^*}$
predictions in~\cite{Altmannshofer:2017fio} that are based on global
fits of $b \to s \mu \mu$ decay data, assuming $b \to s ee$ decays
to be SM-like.

In this letter we interpret the $R_{K^{(*)}}$ measurements using a
model-independent effective Hamiltonian approach
(see~\cite{Hurth:2014vma,Glashow:2014iga,Ghosh:2014awa,Bhattacharya:2014wla,
Alonso:2015sja,Greljo:2015mma,Becirevic:2015asa} for earlier model independent studies of $R_K$).
We also include Belle measurements of LFU observables in the
$B \to K^* \ell^+\ell^-$ angular distibutions~\cite{Wehle:2016yoi}.
We do not consider early results on $R_{K^{(*)}}$ from
BaBar~\cite{Lees:2012tva} and Belle~\cite{Wei:2009zv} which,
due to their large uncertainties, have little impact. We identify the
regions of NP parameter space that give a good description of the
experimental data.
We show how future measurements can lift flat directions in the NP
parameter space and discuss the compatibility of the $R_{K^{(*)}}$
measurements with other anomalies in rare $B$ meson decays.

\paragraph{Model independent implications for new physics.}
%
We assume that NP in the $b \to s \ell\ell$ transitions is sufficiently
heavy such that it can be model-independently described by an effective
Hamiltonian, $\mathcal H_\text{eff} = \mathcal H_\text{eff}^\text{SM} +
\mathcal H_\text{eff}^\text{NP}$,
\begin{equation}
\label{eq:Heff}
\mathcal{H}_\text{eff}^\text{NP} = - \frac{4\,G_F}{\sqrt{2}} V_{tb}V_{ts}^* \frac{e^2}{16\pi^2}
\sum_{i, \ell}
(C_i^\ell O_i^\ell + C_i^{\prime\,\ell} O_i^{\prime\,\ell}) + \text{h.c.} \,,
\end{equation}
with the following four-fermion contact interactions,
\begin{align}
O_9^\ell &=
(\bar{s} \gamma_{\mu} P_{L} b)(\bar{\ell} \gamma^\mu \ell)\,,
& \!\!\!\!\!
O_9^{\prime \, \ell} &=
(\bar{s} \gamma_{\mu} P_{R} b)(\bar{\ell} \gamma^\mu \ell)\,,\label{eq:O9}
\\
O_{10}^\ell &=
(\bar{s} \gamma_{\mu} P_{L} b)( \bar{\ell} \gamma^\mu \gamma_5 \ell)\,,
& \!\!\!\!\!
O_{10}^{\prime \, \ell} &=
(\bar{s} \gamma_{\mu} P_{R} b)( \bar{\ell} \gamma^\mu \gamma_5 \ell)\,,\label{eq:O10}
\end{align}
and the corresponding Wilson coefficients $C_i^\ell$, with $\ell = e,\mu$.
We do not consider other dimension-six operators that can contribute
to $b \to s \ell\ell$ transitions.
Dipole operators and four-quark operators \cite{Jager:2017gal} cannot
lead to violation of LFU and are therefore irrelevant for this work.
Four-fermion contact interactions containing scalar currents would be
a natural source of LFU violation. However, they are strongly constrained
by existing measurements of the $B_s \to \mu\mu$ and $B_s \to ee$
branching ratios~\cite{Aaij:2017vad,Aaltonen:2009vr}. Imposing $SU(2)_L$
invariance, these bounds cannot be avoided~\cite{Alonso:2014csa}. We
have checked explicitly that $SU(2)_L$ invariant scalar operators cannot
lead to any appreciable effects in $R_{K^{(*)}}$ (cf.~\cite{Altmannshofer:2017wqy}).

For the numerical analysis we use the open source code \texttt{flavio}~\cite{flavio}. Based on the experimental measurements and theory predictions for the LFU ratios $R_{K^{(*)}}$ and the LFU differences of $B \to K^* \ell^+\ell^-$ angular observables $D_{P_{4,5}^\prime}$ (see below), we construct a $\chi^2$ function that depends on the Wilson coefficients and that takes into account the correlations between theory uncertainties of different observables.
We use the default theory uncertainties in \texttt{flavio}, in particular
$B\to K^*$ form factors from a combined fit to light-cone sum rule and lattice results \cite{Straub:2015ica}.
The experimental uncertainties are presently dominated by statistics, so their
correlations can be neglected.
For the SM we find $\chi^2_\text{SM} = 24.4$ for 5 degrees of freedom.

\renewcommand{\arraystretch}{1.5}
\begin{table}[tb]
\begin{center}
\input{table_pulls}
\end{center}
\caption{Best-fit values and pulls for scenarios with NP in one
individual Wilson coefficient, taking into account only LFU observables.}
\label{tab:pulls}
\end{table}

\begin{figure}[tbp]
\includegraphics[width=\columnwidth]{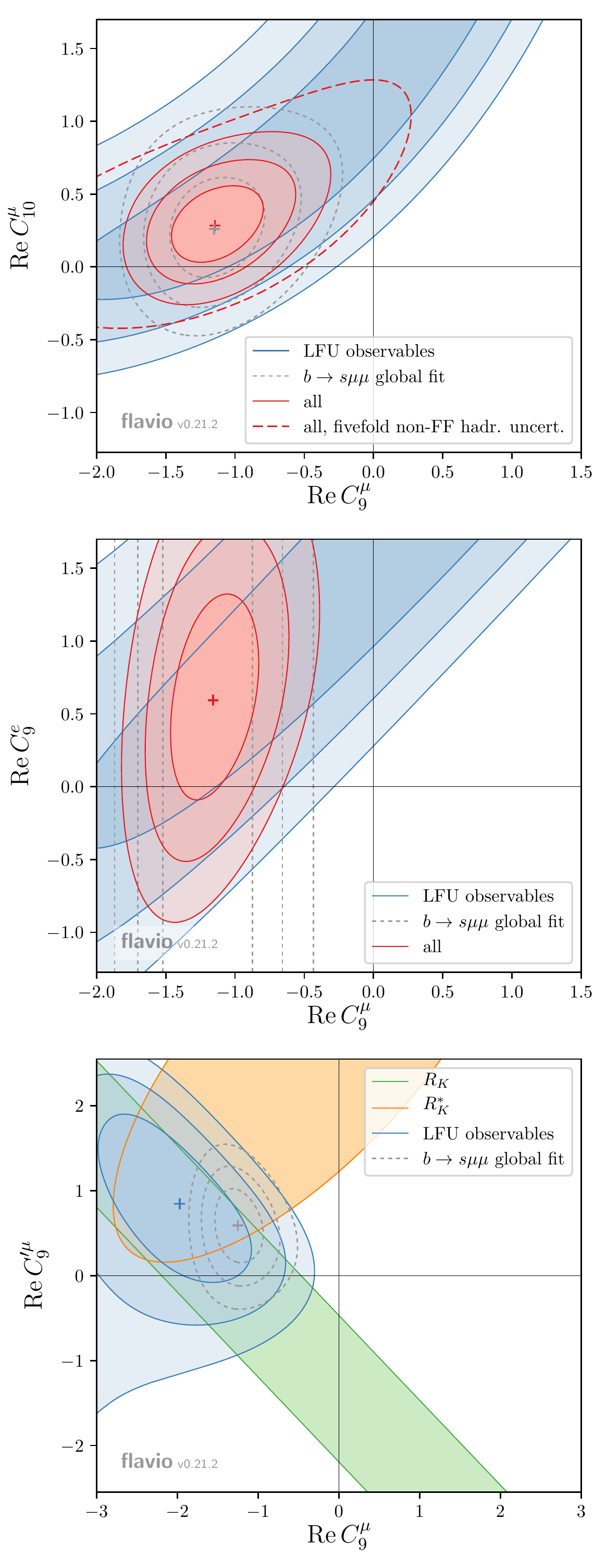}
\caption{Allowed regions in planes of two Wilson coefficients,
assuming the remaining coefficients to be SM-like.}
\label{fig:fits}
\end{figure}

Tab.~\ref{tab:pulls} lists the best fit values and pulls, defined as the
$\sqrt{\Delta \chi^2}$ between the best-fit point and the
SM point for scenarios with NP in one individual Wilson coefficient.
The plots in Fig.~\ref{fig:fits} show contours of constant
$\Delta\chi^2 \approx 2.3, 6.2, 11.8$ in the planes of two Wilson
coefficients for the scenarios with NP in $C_9^\mu$ and $C_{10}^\mu$ (top),
in $C_9^\mu$ and $C_9^e$ (center), or in $C_9^\mu$ and $C_9^{\prime\, \mu}$ (bottom),
assuming the remaining coefficients to be SM-like.

\begin{figure*}[tbp]
\includegraphics[width=\textwidth]{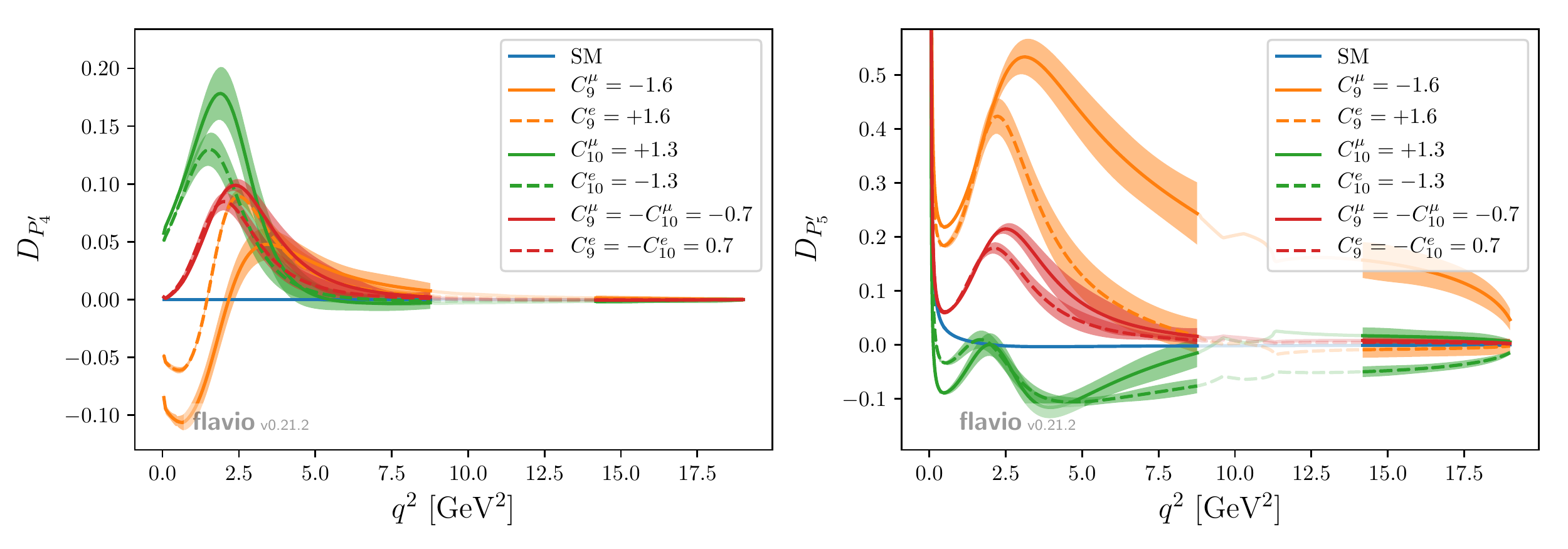}
\caption{The $B \to K^* \ell^+ \ell^-$ LFU differences
$D_{P_4^\prime}$ and $D_{P_5^\prime}$ in the SM and various
NP benchmark models as functions of $q^2$.
The error bands contain all theory uncertainties including
form factors and non-factorisable hadronic effects.
In the region of narrow charmonium resonances, only the
short-distance contribution is shown, without uncertainties.}
\label{fig:DP}
\end{figure*}

The fit prefers NP in the Wilson coefficients corresponding
to left-handed quark currents with high significance $\sim 4\sigma$.
Negative $C_9^\mu$ and positive $C_{10}^\mu$ decrease both
$\mathcal B(B \to K \mu^+\mu^-)$ and $\mathcal B(B \to K^* \mu^+\mu^-)$
while positive $C_9^e$ and negative $C_{10}^e$ increase both
$\mathcal B(B \to K e^+e^-)$ and $\mathcal B(B \to K^* e^+e^-)$,
allowing a good description of the data in each case.
Also along the direction $C_9^\ell = - C_{10}^\ell$ that
corresponds to a left-handed lepton current, we find excellent fits to the data.
As can be seen from the top plot in Fig.~\ref{fig:fits}, right-handed muon
currents ($C_9^\mu = C_{10}^\mu$) cannot describe the data.
We find that right-handed electron currents can explain
$R_K$ and $R_{K^*}$ if the corresponding Wilson coefficients
are sizable ($C_9^e = C_{10}^e \sim -2$ or $+3$).

The primed Wilson coefficients, that correspond to right-handed
quark currents, cannot improve the agreement with the data by
themselves. As is well known~\cite{Hiller:2014ula}, the primed
coefficients imply $R_{K^*} > 1$ given $R_K < 1$ and vice versa.
The complementary sensitivity of $R_{K^*}$ and $R_K$ to right-handed
currents is illustrated in the bottom plot of Fig.~\ref{fig:fits}
for the example of $C_9^\mu$ vs. $C_9^{\prime \,\mu}$.
In combination with sizable un-primed coefficients, the primed coefficients can
slightly improve the fit.

Among the un-primed Wilson coefficients, there are approximate
flat directions. We find that a good description of the
experimental results is given by
\begin{equation}
 C_9^\mu - C_9^e - C_{10}^\mu + C_{10}^e \simeq -1.4 ~,
\end{equation}
unless some of the individual coefficients are much larger than
$1$ in absolute value.
The flat direction is clearly visible in the top and center plot of Fig.~\ref{fig:fits}.
In many NP models one has relations
among these coefficients. In models with leptoquarks one finds
$C_9^\ell = \pm C_{10}^\ell$~\cite{Hiller:2014yaa,Buras:2014fpa}, models based on gauged
$L_\mu - L_\tau$ predict $C_9^e = C_{10}^\ell = 0$~\cite{Altmannshofer:2014cfa},
while in some $Z^\prime$ models one finds $C_9^\ell = a C_{10}^\ell$,
where $a$ is a constant of $\mathcal O(1)$ (see e.g.~\cite{Buras:2016dxz}).

We find that a non-standard $C_{10}^\ell$ ($C_9^\ell$) leads to slightly larger
(smaller) effects in $R_{K^*}$ than in $R_{K}$.
Therefore, $R_{K^*} \lesssim R_K < 1$ is best
described by a non-standard $C_{10}^\ell$. The opposite hierarchy,
$R_{K} \lesssim R_{K^*} < 1$, would lead to a slight preference for
NP in $C_9^\ell$.

A more powerful way to distinguish NP in $C_9^\ell$ from NP in
$C_{10}^\ell$ is through measurements of LFU differences of angular observables~\cite{Altmannshofer:2015mqa,Capdevila:2016ivx,Serra:2016ivr}.
We find that the observables
\begin{align}
D_{P_4^\prime} &= P_4^\prime(B \to K^* \mu^+\mu^-) - P_4^\prime(B \to K^* e^+e^-) \,, \\
D_{P_5^\prime} &= P_5^\prime(B \to K^* \mu^+\mu^-) - P_5^\prime(B \to K^* e^+e^-) \,,
\end{align}
are particularly promising (for a definition of the observables
$P_{4,5}^\prime$ see~\cite{Descotes-Genon:2013vna}).
Predictions for the observables $D_{P_{4,5}^\prime}$ as functions
of $q^2$ in the SM and various NP scenarios are shown in the plots
of Fig.~\ref{fig:DP}. The SM predictions are close to zero with very
high accuracy across a wide $q^2$ range. In the presence of NP,
$D_{P_{4,5}^\prime}$ show a non-trivial $q^2$ dependence.
If the discrepancies in $R_{K^{(*)}}$ are explained by NP in $C_9^\ell$, we predict a negative $D_{P_4^\prime} \sim -0.1$ at low $q^2 \lesssim 2.5$~GeV$^2$ and a sizable positive $D_{P_5^\prime} \sim +0.5$. With NP in $C_{10}^\ell$ we predict instead a positive $D_{P_4^\prime} \sim + 0.15$ and a small negative $D_{P_5^\prime} \sim -0.1$.
We observe that $D_{P_5^\prime}$ has even the potential to distinguish between NP in $C_9^e$ and $C_9^\mu$.
For $q^2 \gtrsim 5$~GeV$^2$, a negative $C_9^\mu$ leads to a sizable increase of $P_5^\prime(B \to K^* \mu^+\mu^-)$, while a positive $C_9^e$ can decrease $P_5^\prime(B \to K^* e^+e^-)$ only slightly, as the SM prediction for $P_5^\prime$ in this $q^2$ region is already close to its model-independent lower bound of $-1$.
The recent measurements by Belle, $D_{P_4^\prime}^{[1,6]} = +0.498 \pm 0.553$ and $D_{P_5^\prime}^{[1,6]} = +0.656 \pm 0.496$~\cite{Wehle:2016yoi}, have still sizable uncertainties and are compatible with NP both in $C_9^\ell$ and in $C_{10}^\ell$. They slightly favor NP in $C_9^\ell$.
We note that, while the SM prediction for these observables has a tiny uncertainty,
for fixed values of LFU violating Wilson coefficients, form factor and other
hadronic uncertainties do play a role, as also shown in Fig.~\ref{fig:DP}.
However, these uncertainties are still so small that sufficient experimental precision could allow a clean identification of the underlying NP contact interaction.

\begin{figure*}[tbp]
\includegraphics[width=\textwidth]{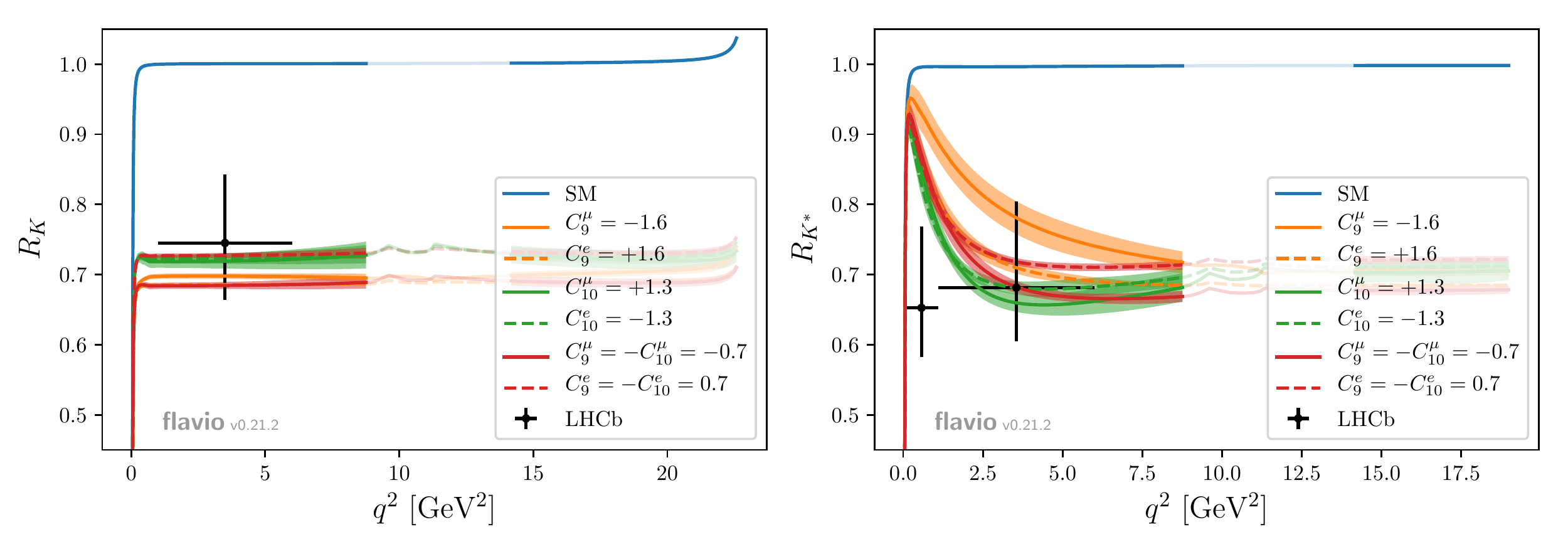}
\caption{The LFU ratios $R_{K^{(*)}}$ in the SM and two
NP benchmark models as function of $q^2$.
Concerning the error bands, the same comments as for Fig.~\ref{fig:DP} apply.}
\label{fig:q2}
\end{figure*}

We stress that the NP contact interactions in~(\ref{eq:Heff}) lead
also to a characteristic $q^2$ shape in the LFU ratios $R_{K^{(*)}}$.
In Fig.~\ref{fig:q2} we show $R_{K^{(*)}}$ as functions of $q^2$ in
the SM and in the same NP scenarios as in Fig.~\ref{fig:DP}. In the SM, $R_{K^{(*)}}$ are to an excellent
approximation $q^2$ independent. For very low $q^2 \simeq 4m_\mu^2$
they drop to zero, due to phase space effects.
NP contact interactions lead to an approximately constant shift in $R_K$.
The ratio $R_{K^*}$, on the other hand, shows a non-trivial $q^2$
dependence in the presence of NP. In contrast to $B \to K \ell \ell$,
the $B \to K^* \ell \ell$ decays at low $q^2$ are dominated by the
photon pole, which gives a lepton flavor universal contribution. The
effect of NP is therefore diluted at low $q^2$.
Given the current experimental uncertainties, the measured $q^2$ shape
of $R_{K^*}$ is compatible with NP in form of a contact interaction.
Significant discrepancies from the shapes shown in Fig.~\ref{fig:q2}
would imply the existence of light NP degrees of freedom around or
below the scale set by $q^2$ and a breakdown of the effective Hamiltonian
framework.

Assuming that the description in terms of contact interactions holds, we translate the best fit values of the Wilson coefficients into a generic NP scale. Reparameterizing the effective Hamiltonian~(\ref{eq:Heff}) as $\mathcal H_\text{eff}^\text{NP} = - \sum_i \mathcal O_i / \Lambda_i^2$, one gets
\begin{equation}
 \Lambda_i = \frac{4\pi}{e} \frac{1}{\sqrt{|V_{tb} V_{ts}^*|}} \frac{1}{\sqrt{|C_i|}} \frac{v}{\sqrt{2}} \simeq \frac{35~\text{TeV}}{\sqrt{|C_i|}}~.
\end{equation}
Based on perturbative unitarity we therefore predict the existence of NP degrees of freedom below a scale of $\Lambda_\text{NP} \sim \sqrt{4\pi} \times 35~\text{TeV} / \sqrt{|C_i|} \sim 100~\text{TeV}$.

\paragraph{Compatibility with other rare \texorpdfstring{$B$}{B} decay anomalies.}

It is natural to connect the discrepancies in $R_{K^{(*)}}$ to the other existing anomalies in rare decays based on the  $b \to s \mu \mu$ transition. In the plots of Fig.~\ref{fig:fits} we show in dotted gray the 1, 2, and 3$\sigma$ contours from our global $b \to s \mu \mu$ fit that does \textit{not} take into account the measurements of the LFU observables $R_{K^{(*)}}$ and $D_{P_{4,5}'}$~\cite{Altmannshofer:2017fio}.
We observe that the blue regions prefered by the LFU observables are fully compatible with the $b \to s \mu \mu$ fit.
We have also performed a full fit, taking into account all the observables from the
$b \to s \mu \mu$ fit, the branching ratio of $B_s\to\mu^+\mu^-$ (assuming it not to be affected by scalar NP contributions),
and the BaBar measurement of the $B\to X_se^+e^-$ branching ratio \cite{Lees:2013nxa}.
This fit, shown in red, points to a non-standard $C_9^\mu \simeq - 1.2$ with very high significance.
Wilson coefficients other than $C_9^\mu$ are constrained by the global fit.

Compared to the LFU observables, the global $b \to s \mu \mu$ fit depends more strongly on estimates of hadronic uncertainties in the $b \to s \ell \ell$ transitions.
To illustrate the impact of a hypothetical, drastic underestimation of these uncertainties,
we also show results of a global fit where uncertainties of non-factorisable hadronic contributions are inflated by a factor of 5 with respect to our nominal estimates. In this case, the global fit becomes dominated
by the LFU observables, but the $b\to s\mu\mu$ observables still lead to relevant
constraints. For instance, the best-fit value for $C_{10}^\mu$ in Tab.~\ref{tab:pulls}
would imply a 50\% suppresion of the $B_s\to\mu^+\mu^-$ branching ratio, which
is already in tension with current measurements \cite{Aaij:2017vad},
barring cancellations with scalar NP contributions.

\paragraph{Conclusions.}

The discrepancies between SM predictions and experimental results in the LFU ratios $R_K$ and $R_{K^*}$ can be explained by NP four-fermion contact interactions $(\bar s b)(\bar \ell \ell)$ with left-handed quark currents. Future measurements of LFU differences of $B \to K^* \ell^+\ell^-$ angular observables can help to identify the chirality structure of the lepton currents.
If the hints for LFU violation in rare $B$ decays are first signs of NP, perturbative unitarity implies new degrees of freedom below a scale of $\Lambda_\text{NP} \sim 100$~TeV.
These results are robust, i.e. they depend very mildly on assumptions about the size of hadronic uncertainties in the $B\to K^{(*)} \ell^+ \ell^-$ decays.

Intriguingly, the measured values of $R_K$ and $R_{K^*}$ are fully compatible with NP explanations of various additional anomalies in rare $B$ meson decays based on the $b \to s \mu\mu$ transition. A combined fit singles out NP in the Wilson coefficient $C_9^\mu$ as a possible explanation.

\subsection*{Acknowledgments}

WA acknowledges financial support by the University of Cincinnati.
The work of PS and DS was supported by the DFG cluster of
excellence ``Origin and Structure of the Universe''.

\bibliography{bibliography}

\end{document}

%% file: table_pulls.tex
\begin{tabularx}{0.48\textwidth}{ccccX}
\hline\hline
 ~~~ Coeff. ~~~ & ~best fit~ & $1\sigma$ & $2\sigma$ & pull\\
\hline\hline
\rowcolor[gray]{.9} $C_9^{\mu}                      $ & $-1.56$ & [$-2.12$, $-1.10$] & [$-2.87$, $-0.71$] & $4.1\sigma$\\
                    $C_{10}^{\mu}                   $ & $+1.20$ & [$+0.88$, $+1.57$] & [$+0.58$, $+2.00$] & $4.2\sigma$\\
\rowcolor[gray]{.9} $C_9^{e}                        $ & $+1.54$ & [$+1.13$, $+1.98$] & [$+0.76$, $+2.48$] & $4.3\sigma$\\
                    $C_{10}^{e}                     $ & $-1.27$ & [$-1.65$, $-0.92$] & [$-2.08$, $-0.61$] & $4.3\sigma$\\
\rowcolor[gray]{.9} $C_9^{\mu}=-C_{10}^{\mu}        $ & $-0.63$ & [$-0.80$, $-0.47$] & [$-0.98$, $-0.32$] & $4.2\sigma$\\
                    $C_9^{e}=-C_{10}^{e}            $ & $+0.76$ & [$+0.55$, $+1.00$] & [$+0.36$, $+1.27$] & $4.3\sigma$\\
\rowcolor[gray]{.9} $C_9^{e}=C_{10}^{e}             $ & $-1.91$ & [$-2.30$, $-1.51$] & [$-2.71$, $-1.10$] & $3.9\sigma$\\
\hline
                    $C_9^{\prime\,\mu}              $ & $-0.05$ & [$-0.31$, $+0.21$] & [$-0.57$, $+0.46$] & $0.2\sigma$\\
\rowcolor[gray]{.9} $C_{10}^{\prime\,\mu}           $ & $+0.03$ & [$-0.21$, $+0.27$] & [$-0.44$, $+0.51$] & $0.1\sigma$\\
                    $C_9^{\prime\,e}                $ & $+0.07$ & [$-0.21$, $+0.37$] & [$-0.49$, $+0.69$] & $0.2\sigma$\\
\rowcolor[gray]{.9} $C_{10}^{\prime\,e}             $ & $-0.04$ & [$-0.30$, $+0.21$] & [$-0.57$, $+0.45$] & $0.2\sigma$\\
\hline\hline
\end{tabularx}